\documentclass[twocolumn,superscriptaddress,prl]{revtex4}
\usepackage{graphicx}
\usepackage{color}
\usepackage{epsfig,subfigure}
\usepackage{amsmath,amssymb,latexsym}

\def\U#1{{\rm #1}} 
\def\u#1{_{\rm #1}}
\newcommand{\bra}[1]{\langle #1 |}
\newcommand{\ket}[1]{| #1 \rangle}
\newcommand{\ketbra}[2]{| #1 \rangle \langle #2 |}
\newcommand{\expect}[1]{\left\langle #1 \right\rangle} 
\newcommand{\dagg}[1]{#1 ^\dagger}
\def\tr{\U{tr}}
\def\DRD{\U{\U{D}\u{RD}}}
\def\DRA{\U{\U{D}\u{R\bar{D}}}}
\def\DLD{\U{\U{D}\u{LD}}}
\def\DLA{\U{\U{D}\u{L\bar{D}}}}
\def\vac{{\rm vac}}
\def\prl{Phys. Rev. Lett.}
\def\pra{Phys. Rev. A}

\begin{document}
\title{
Measurement-device-independent quantum key distribution 
for Scarani-Acin-Ribordy-Gisin 04 protocol
}
\author{Akihiro Mizutani}
\affiliation{Graduate School of Engineering Science, Osaka University,
Toyonaka, Osaka 560-8531, Japan}
\author{Kiyoshi Tamaki}
\affiliation{NTT Basic Research Laboratories, NTT Corporation, 
3-1, Morinosato-Wakamiya Atsugi-Shi, 243-0198, Japan\\
${}^*$imoto@mp.es.osaka-u.ac.jp}
\author{Rikizo Ikuta}
\affiliation{Graduate School of Engineering Science, Osaka University, Toyonaka, Osaka 560-8531, Japan}
\author{Takashi Yamamoto}
\affiliation{Graduate School of Engineering Science, Osaka University, Toyonaka, Osaka 560-8531, Japan}
\author{Nobuyuki Imoto$^*$}
\affiliation{Graduate School of Engineering Science, Osaka University, Toyonaka, Osaka
560-8531, Japan}
\begin{abstract}
The measurement-device-independent quantum key distribution~(MDI QKD) 
was proposed to make BB84 completely free from any side-channel in detectors. 
Like in prepare \& measure QKD, 
the use of other protocols in MDI setting would be advantageous 
in some practical situations. 
In this paper, we consider SARG04 protocol in MDI setting. 
The prepare \& measure SARG04 is proven to be able to generate a key 
up to two-photon emission events. 
In MDI setting we show that the key generation is possible 
from the event with single or two-photon emission 
by a party and single-photon emission by the other party, 
but the two-photon emission event by both parties cannot contribute 
to the key generation. On the contrary to prepare \& measure SARG04 protocol 
where the experimental setup is exactly the same as BB84, 
the measurement setup for SARG04 in MDI setting cannot be the same as that 
for BB84 since the measurement setup for BB84 in MDI setting induces 
too many bit errors. To overcome this problem, 
we propose two alternative experimental setups, 
and we simulate the resulting key rate. 
Our study highlights the requirements that MDI QKD poses on us regarding 
with the implementation of a variety of QKD protocols. 
\end{abstract}

\maketitle

The security of quantum key distribution~(QKD) 
can be guaranteed based on some mathematical models 
of the users' devices~\cite{QKD, QKD2, QKD3}. 
Unfortunately, the actual devices 
do not necessarily follow mathematical models, 
and we need to close the gap~(side-channel) 
between the actual device and the mathematical model 
to implement secure QKD systems in practice. 
Among side-channels, 
the side-channel of a photon detector seems to be most easily 
exploited by an eavesdropper~(Eve) since it accepts any input from Eve who can
generate an arbitrary optical state such that it causes an unexpected behavior in the
detector. In fact, the famous bright-pulse illumination attacks are 
based on side-channel in detectors~\cite{Makarov}. 
In order to countermeasure such attacks, measurement-device-independent~(MDI) 
QKD~\cite{MDIQKD} was proposed to make BB84~\cite{BB84} 
free from any possible side-channel in a detector. 
In MDI QKD, Alice and Bob do not perform any measurement 
but only send quantum signals to be measured by Eve. 
Therefore, bit strings generated by Alice and Bob are 
free from side-channels in photon detectors since they do not 
employ photon detectors. Since its invention, 
MDI QKD has been actively studied 
both theoretically~\cite{MDIBB84th1,MDIBB84th2} and experimentally~\cite{MDIBB84ex1,MDIBB84ex2}. 

As is the case in prepare \& measure scheme, 
implementation of protocols other than BB84 in MDI setting could be suitable 
for some practical situations. 
In fact, many experiments for non-BB84 type prepare \& measure schemes, 
including B92~\cite{B92}, DPS QKD~\cite{DPS}, 
coherent one-way protocol~\cite{Gisin}, 
SARG04~\cite{SARG04}, etc, have been reported~\cite{Sasaki}. 
Therefore, it is useful in practice 
to use non-BB84 type protocols in MDI setting, 
and in this paper we consider to use SARG04 protocol in MDI setting, 
which we refer to as MDI SARG04. 
SARG04 was originally proposed to make BB84 
robust against photon number splitting (PNS) attacks~\cite{PNS1,PNS2} just 
by changing the classical post-processing part in BB84. 
It is proven that SARG04 can indeed generate a key 
from two-photon emission event by Alice in addition to 
single-photon emission event~\cite{SARGtamaki,SARGkoashi}, 
showing robustness against PNS attack in some parameter regimes. 
Note in MDI setting is that both 
Alice and Bob are the sender of the signals, 
and as a result, the information leakage from the signals 
seems to be larger than the one in prepare \& measure setting. 
Therefore, it is not trivial whether both single and two-photon emission events 
can contribute to the key generation or not. 
Our work answers this question, 
and we have found that the single-photon emission event 
by both Alice and Bob, 
or single-photon and two-photon emission by each of Alice and Bob can 
contribute to generating a key, 
but two-photon emission by the two parties cannot make the contribution 
when a probability of Eve's announcement 
of the successful measurement 
for the two-photon emission event is smaller than 1/16. 

Another important issue to be addressed in MDI setting is 
what kind of measurement setup should be implemented experimentally 
at Eve's laboratory. 
Naively thinking, as SARG04 differs from BB84 only in the post-processing part, 
the same measurement setup for MDI BB84 should also work for MDI-SARG04 protocol. 
On the contrary, 
however, it turns out that the measurement setup for MDI BB84 results 
in high bit error rate when applied to MDI-SARG04 protocol, 
and consequently, no significant key can be generated. 
To generate a key in practice, 
we propose two alternative measurement 
schemes for the MDI-SARG04 protocol, 
and simulate the resulting key generation rate. 

\section{Results}
\subsection{MDI-SARG04 QKD protocol}
\label{sec:protocol}
\begin{figure}[t] 
 \begin{center}
 \includegraphics[width=7cm,clip]{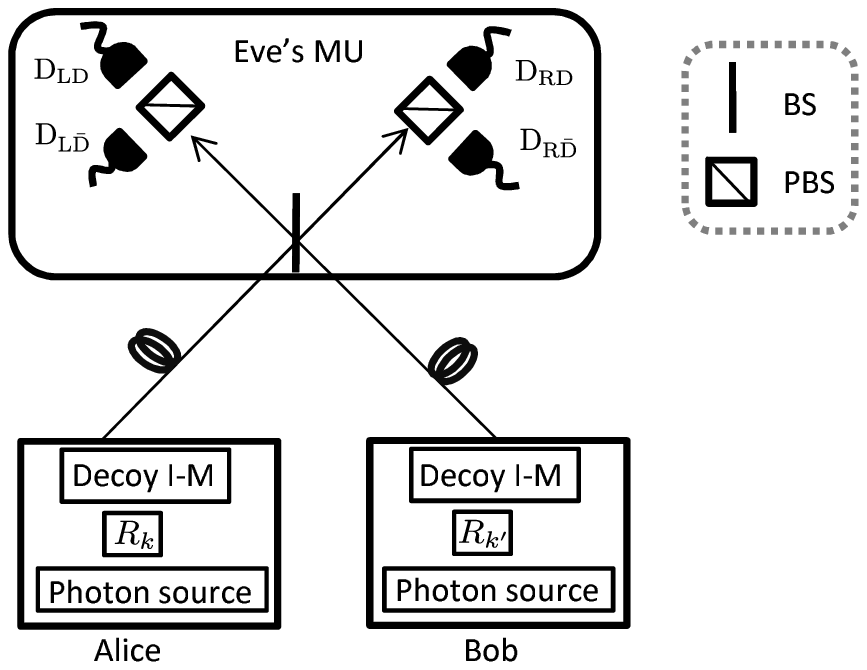}
\end{center}
\caption{
Schematic of an experimental setup 
for the MDI-SARG04 QKD. 
The role of Eve's measurement unit~(MU) 
is to perform entangling operation on the photons 
from Alice and Bob, which is implemented by using a half beamsplitter~(BS)
followed by polarization BSs~(PBSs) and photon detectors. 
We note that the PBS passes the photons in $45^\circ$ polarization 
and reflects the photons in $-45^\circ$ polarization. 
}
\label{fig:actual}
\end{figure}

In this section, 
we introduce the MDI-SARG04 QKD protocol. 
First, we summarize the assumptions and mathematical definitions 
made in this paper, 
and then we describe how the protocol runs. 

\subsubsection{Assumptions and definitions}
\label{sec:protocol1}
We assume that each of Alice and Bob has a phase randomized photon 
source, {\it i.e.} the vacuum, a single photon, and multi photons are 
emitted probabilistically. 
The probabilities of the $n$-photon emission from Alice and Bob 
are $p_n$ and $p_{n'}$, respectively, 
which satisfy $\sum_n p_n =\sum_{n'} p_{n'}=1$.  
We encode the bit information in polarization of photons, 
and we assume that the preparation of the polarization is precise without any flaw. 
For simplicity, we consider the asymptotic case 
to neglect any statistical fluctuation, i.e., 
the number of the signals sent by Alice and Bob is infinite. 
In our paper, 
horizontal and vertical polarization states of a single photon 
are represented by $Z$-basis qubit states, namely 
$\ket{0_{z}}$ and $\ket{1_{z}}$, respectively. 
We also define $X$~(rectilinear)-basis states as 
$\ket{i_{x}}=(\ket{0_{z}}+(-1)^{i}\ket{1_{z}})/\sqrt{2}$ for $i=0,1$. 
By using a creation operator $\dagg{a}_{\theta}$ for a single photon 
in a polarization $\theta$ and the vacuum state $\ket{\vac}$, 
we denote an $n$-photon number state with polarization $\theta$ 
by $\ket{n_{\theta}}=(\dagg{a}_{\theta})^n\ket{\vac}/\sqrt{n!}$.
~(note that when the subscript $\theta$ is $z$ or $x$, 
it refers to the qubit state rather than the photon number state). 
Other definitions we use are as follows: 
$\ket{\varphi_{i}}=\cos(\pi/8)\ket{0_{x}}
+(-1)^{i}\sin(\pi/8)\ket{1_{x}}$ for $i=0,1$ and 
$\ket{\varphi_{i}}=\sin(\pi/8)\ket{0_{x}}
+(-1)^{i-1}\cos(\pi/8)\ket{1_{x}}$ for $i=2,3$. 
$R=\exp(-\pi/2 Y)$, where $Y=-i\ketbra{0_z}{1_z}+i\ketbra{1_z}{0_z}$, 
which satisfies 
$R\dagg{a}_{\varphi_i}\dagg{R}=\dagg{a}_{\varphi_{i+1(\U{mod} 4)}}$ for all $i$. 
$\ket{\psi^{\pm}}=(\ket{0_x1_x}\pm\ket{1_x0_x})/\sqrt{2}$ and 
$\ket{\phi^+}=(\ket{0_x0_x}+\ket{1_x1_x})/\sqrt{2}$. 
We denote ${P}(\cdot) = {(\cdot)(\cdot)^{\dag}}$. 

\subsubsection{The protocol of the MDI-SARG04 QKD}
\label{sec:protocol2}
The protocol runs as follows:  

(a1)
Alice and Bob choose a bit value $i$ and $i'$~$(i,i'=0,1)$, 
respectively, and they encode the bit value into 
the photonic states of their pulses as 
$\sum_n p_n\ketbra{n_{\varphi_i}}{n_{\varphi_i}}$
and 
$\sum_{n'} p_{n'}\ketbra{n'_{\varphi_{i'}}}{n'_{\varphi_{i'}}}$. 

(a2)
Alice and Bob rotate the polarization of their pulses 
by applying rotation $R_k$ and $R_{k'}$ 
with randomly-chosen values of $k (=0,1,2,3)$ and $k' (=0,1,2,3)$, 
respectively, where $R_k$ is defined by $R_k\equiv R^k$. 
After the rotation, 
Alice and Bob send the pulses to 
Eve's measurement unit~(MU) through quantum channels. 

(a3)
Eve performs a measurement on the incoming pulses 
and announces to Alice and Bob over the authenticated public channel 
whether her measurement outcome is successful or not. 
When the outcome is successful, 
she also announces types of the successful events, 
either Type1 or Type2. 

(a4)
Alice and Bob broadcast $k$ and $k'$, 
over the authenticated public channel. 
If the measurement outcome in (a3) is successful with Type1 
and $k=k'=0,\ldots,3$, 
they keep their bit values $i$ and $i'$ in (a1), and Alice flips her bit. 
If the measurement outcome in (a3) is successful with Type2 
and $k=k'=0, 2$, they keep their bit values $i$ and $i'$ in (a1). 
In all the other cases, they discard their bit values. 

\begin{table}[t]
\begin{center}
\begin{tabular}
{|c|c|c|}
\hline
 successful event & output \\ \hline \hline
Type1 ($\DLD$\&$\DRA$ or $\DRD$\&$\DLA$) &$\ket{\psi^-}$ \\ \hline
Type2 ($\DLD$\&$\DLA$ or $\DRD$\&$\DRA$) &$\ket{\psi^+}$ \\ \hline
\end{tabular}
\end{center}
\caption{Two types of the successful events announced by 
Eve's MU. 
Type1 is the coincidence detection events 
of $\DLD$\&$\DRA$ or $\DRD$\&$\DLA$ denoted in  Fig.~\ref{fig:actual}. 
Type2 is the coincidence detection events 
of $\DLD$\&$\DLA$ or $\DRD$\&$\DRA$. 
When the successful events are Type1 and Type2, 
Alice and Bob distill the states $\ket{\psi^-}$ and $\ket{\psi^+}$, respectively, 
in the virtual protocol. 
 \label{tbl:type}}
\end{table}

(a5)
Alice and Bob repeat from (a1) to (a4) 
until 
the number of 
the successful events with rotation $k=k'=0,\ldots, 3$ in Type1 becomes $N_1$ 
and $k=k'=0, 2$ in Type2 becomes $N_2$. 
Let $N_iQ_{i}^\U{tot}$ be the number of the successful detection event of Type $i$. 
Alice and Bob announce randomly-chosen $N_iQ_{i}^\U{tot}\zeta$ bits 
over the authenticated public channel, 
where $\zeta$ is much smaller than 1, 
and estimate the error rate $e_{i}^\U{tot}$ in the remaining code bits. 
The estimated number of the bit error in the code bits is denoted by 
$e_{i}^\U{tot}N_iQ_{i}^\U{tot}(1-\zeta)$. 

(a6) 
Alice and Bob perform error correction and privacy amplification 
on the remaining $N_iQ_i^\U{tot}(1-\zeta)$ bits 
by their discussion over the public channel. 
As a result, they share a final key of length 
$G_1N_1(1-\zeta)+G_2N_2(1-\zeta)$. 

At Eve's MU in (a3), 
honest Eve performs the Bell measurement 
in order to establish quantum correlations 
between Alice and Bob to generate the key. 
In Fig.~\ref{fig:actual}, 
the experimental setup for the Bell measurement is depicted. 
It employs a half beam splitter~(BS), 
two polarization BSs~(PBSs), 
and the photon detectors. 
In the case where both Alice and Bob emit a single photon, 
the simultaneous photon detection events matching 
the pattern Type1~(Type2), listed in Table~\ref{tbl:type}, 
corresponds to the detection of $\ket{\psi^-}$~($\ket{\psi^+}$). 
We emphasize that in the security proof we assume 
that Eve is malicious and has a control over the quantum channels, 
and all the bit errors are attributed to the consequence of the eavesdropping. 

\subsection{Limitation of the experimental setup}
\label{sec:limitation}
In prepare \& measure setting, the SARG04 protocol is different 
from the BB84 protocol only in the post-processing part, i.e., 
no modification is needed in the experimental setup. 
In the MDI setting, however, the experimental setup 
for the BB84 protocols~\cite{MDIQKD} cannot be directly used in MDI-SARG04 
as it induces a high bit error rate, 
and this is a significant qualitative difference of MDI setting 
from prepare \& measure setting, 
implying that not all the prepare \& measure QKD protocols 
cannot be directly converted to MDI setting. 
Therefore, we need to consider an alternative experimental scheme for MDI-SARG04. 
In this section, we first discuss why the setup for MDI-BB84 gives 
the high bit error rate, 
and then we propose alternative experimental schemes for MDI-SARG04. 

For the explanation we denote by $F^{(n,m)}$ 
the joint probability that Eve receives $n$ and $m$ photons from Alice and Bob, 
respectively, and obtains the successful measurement outcome. 
Note that while we do not deal with the types of 
Eve's successful outcomes separately, 
the following discussion is valid for both types. 
For simplicity, we neglect all the losses, 
including those in the quantum channel and the photon detectors, 
and therefore we can also regard $F^{(n,m)}$ as $Q^{(n,m)}$, 
which is the joint probability that Alice and Bob respectively 
emit $n$ and $m$ photons and Eve obtains the successful measurement outcome. 
Like in the MDI-BB84 protocols, we assume that 
Alice and Bob use a phase randomized weak coherent light 
whose average photon number is much smaller than 1. 
Thus, we have 
$Q^{(1,1)}/2\sim Q^{(2,0)}\sim Q^{(0,2)} \gg Q^{(n,m)}$ for $n+m\geq 3$. 
For simplicity, we assume Eve is honest, namely 
the bit error rate for $n=m=1$ is zero, 
and all photon detectors have unit quantum efficiency and no dark counting. 
In the following, 
we show that even with this simplification favorable to Alice and Bob, 
no significant key is expected. 
To see this, we consider the bit error rate, 
and the total bit error rate $e^\U{tot}$ is expected to be 
\begin{eqnarray}
e^\U{tot} \sim 
\frac{Q^{(2,0)}e\u{bit}^{(2,0)}+Q^{(0,2)}e\u{bit}^{(0,2)}}
{Q^{(1,1)}+Q^{(2,0)}+Q^{(0,2)}}, 
\label{eq:etotapprox}
\end{eqnarray}
where $e\u{bit}^{(n,m)}$ is the bit error probability under the condition 
that Alice emits $n$ photons and Bob emits $m$ photons, 
and Eve announces the successful outcome. 
Note that equation~(\ref{eq:etotapprox}) holds 
in both the MDI-BB84 and MDI-SARG04 protocols. 
It is clear from equation~(\ref{eq:etotapprox}) that 
the bit error is caused by the case 
where one party emits two photons and the other party emits the vacuum. 
It is also clear that $e\u{bit}^{(2,0)}$ cannot be zero 
since the vacuum emission carries no bit information. 
In the case of MDI-BB84, 
this event is always discarded from the sifted key, 
and consequently the bit error rate in the key generation basis, 
i.e., rectilinear basis, is zero. 
This is so because the two-photon states 
$\ket{2_{45^\circ}}$ and $\ket{2_{-45^\circ}}$, which contribute to the bit values, 
are orthogonal and they never produce the successful outcomes 
in Eve's projection measurement for the basis $\{ \ket{0_x}, \ket{1_x}\}$. 
Therefore, in the experiment of MDI-BB84, 
the bit error rate is very small. 
In the case of MDI-SARG04, however, two states 
$\ket{2_{\varphi_0}}$ and $\ket{2_{\varphi_1}}$ 
consisting bit values are not orthogonal. 
This means that the two-photon emission 
contributes to the successful outcome. More precisely, 
$e^\U{tot} \sim e\u{bit}^{(0,2)}/2=0.25$ holds 
from the direct calculation of $e\u{bit}^{(2,0)}=e\u{bit}^{(0,2)}=0.5$
\footnote{
Note that $Q^{(1,1)}/2\sim Q^{(2,0)}\sim Q^{(0,2)}$ and $e^\U{tot} \sim 0.25$ hold 
for any linear loss transmittance channel.}.
Therefore, 
we conclude that the use of the 
phase randomized coherent light source gives no significant key in MDI-SARG04. 
\begin{figure}[t] 
\begin{center}
 \includegraphics[width=8cm,clip]{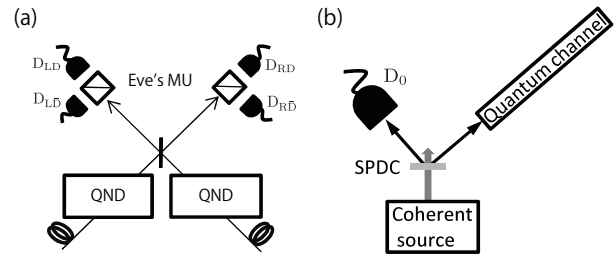}
\end{center}
\caption{
Two experimental setups for generating the key 
in the MDI-SARG04 protocol. 
Both setups significantly eliminate the events caused by $(n,m)=(2,0)$, $(0,2)$ 
and other problematic photon number configurations. 
(a) 
Eve performs the QND measurements on the pulses from Alice and Bob, 
and she does not perform the interference measurement for $n\geq 2$ or $m\geq 2$. 
Eve accepts only when $n\leq 1$ and $m\leq 1$ are satisfied. 
(b)
A quasi single-photon source used by Alice and Bob, 
which is composed of the heralded SPDC  process. 
When detector $\U{D}_0$ clicks, 
Alice/Bob sends her/his pulse at the remaining mode to Eve's MU. 
}
\label{fig:qnd+parame}
\end{figure}
In order to generate a key in the MDI-SARG04 protocol, 
Eve's MU or the photon sources should be modified 
such that the probability of obtaining the successful outcome 
due to the two photons and the vacuum state is suppressed. 
In order to suppress the probability, 
we propose two experimental setups: 
(i) 
Eve performs quantum nondemolition~(QND) measurement 
on the two incoming pulses from Alice and Bob just before mixing them 
as shown in Fig.~\ref{fig:qnd+parame}(a). 
The QND measurement discriminates 
whether the photon number in the pulse is 0, 1 or more. 
Eve accepts only the case where $n\leq 1$ and $m\leq 1$
and discards the other cases with multiple photons. 
Thanks to the QND measurement, the total bit error rate is 
suppressed even if the phase randomized coherent light is used as a photon source. 
(ii)
Without the modification of Eve's MU, 
Alice and Bob replace the phase randomized coherent light 
by a heralded single photon source 
based on a spontaneous parametric down-conversion~(SPDC) 
and a threshold photon detector~(see Fig.~\ref{fig:qnd+parame}(b)). 
This dramatically reduces the probabilities 
of the events of $(n,m)=(2,0)$ and $(0,2)$. 
We will show 
that these setups enable us to generate the key later. 

\subsection{Security Proof}
\label{sec:securityproof}
In this section, 
we discuss the unconditional security proof~(i.e., 
the security proof against most general attacks) of our scheme. 
The security proof is independent of 
the specific device models like in Fig.~\ref{fig:qnd+parame}, 
namely it is valid for any Eve's MU 
and any photon sources of Alice and Bob. 
Our proof employs the security proof based on 
the entanglement distillation protocol~(EDP)~\cite{QKD3, EDP}, 
where the distillation of 
$\ket{\psi^{-}}$ is considered for Type1 and 
that of 
$\ket{\psi^{+}}$ is considered for Type2. 
The proposed EDP-based virtual protocol, 
which is equivalent to the MDI-SARG04 QKD from Eve's viewpoint, 
runs as follows. 

(v1)
Alice and Bob prepare 
$\ket{\Phi_{n(m),k(k')}}\u{A_1(B_1),A_2(B_2)}$, 
where 
$\ket{\Phi_{n,k}}\u{\Gamma_1,\Gamma_2}
=(\ket{0_z}\u{\Gamma_1}\ket{n_{\varphi_{k}}}\u{\Gamma_2}
+\ket{1_z}\u{\Gamma_1}\ket{n_{\varphi_{1+k}}}\u{\Gamma_2})/\sqrt{2}$ 
for $\Gamma$=A,B. 
Here $k(=0,1,2,3)$ and $k'(=0,1,2,3)$ are randomly chosen. 
The probability distribution of the photon number is 
equal to that of 
the photon source in the actual protocol. 
Alice and Bob send the $n$ and $m$ photon states 
in $\U{A_2}$ and $\U{B_2}$ to Eve's MU, respectively. 

(v2)
Eve performs a measurement on the photons coming from Alice and Bob, 
and announces to them 
whether the measurement is 
successful~(including the type of the event) or not. 
If the measurement result is not successful, 
Alice and Bob discard their qubits. 

(v3)
Alice and Bob broadcast the labels $k$ and $k'$, respectively. 
In the cases of $k=k'=1,3$ with the announcement of Type2 
or $k\neq k'$, Alice and Bob discard their qubits.

(v4)
Alice and Bob repeat (v1) -- (v3) many times 
until the number of the successful events for $k=k'$ becomes $N_i$ 
for $i=1,2$, where $i$ corresponds to the type of the events. 

(v5)
Let $N_iQ_i^\U{tot}$ be the number of the successful detection event 
for Type $i$. 
Alice and Bob announce randomly chosen $N_iQ_{i}^\U{tot}\zeta$-photon pairs 
over the authenticated public channel, where $\zeta$ is much smaller than 1, 
and then they perform $Z$-basis measurement on their qubits of the chosen pairs. 
By sharing their measurement results over the authenticated public channel, 
they estimate the bit error rate on the code qubits denoted by $e_i^\U{tot}$. 
As a result, the number of the bit error is 
estimated to be $e_i^\U{tot}N_iQ_i^\U{tot}(1-\zeta)$. 

(v6)
They estimate the upper bound on the phase error rate 
$e^{(n,m)}_{i,\U{ph}}$ for $n$ and $m$ photons 
from the bit error rate $e^{(n,m)}_{i,\U{bit}}$ for $n$ and $m$ photons. 
Here the phase error is defined by the bit error 
that would have been obtained 
if they had measured the qubit pairs by $X$ basis, 
which is the complementarity basis of the computational basis. 

(v7)
When the bit and the phase errors are smaller than a threshold value 
for entanglement distillation, they perform the distillation 
for $N_iQ_i^\U{tot}(1-\zeta)$ qubit pairs. 
For the cases of Type1 and Type2, 
they distill the photon pairs 
in states $\ket{\psi^-}$ and $\ket{\psi^+}$, respectively. 
We denote the number of the distilled maximally 
entangled qubit pairs as $G_iN_i(1-\zeta)$.  
Finally, by performing 
$Z$-measurements on the distilled photon pairs, 
they obtain the key. 

The important quantities in the proof is the bit and phase errors, 
and the phase error rate determines the amount of privacy amplification. 
The bit error rate in the code bits of the virtual protocol, 
which is exactly the same as the one of the actual protocol, 
is directly estimated by test bits. On the other hand, 
the phase error rate is defined by the complementary basis $X$, 
which Alice and Bob never employ, 
and therefore this rate is not directly estimated in the protocols. 
Note that we are allowed to work on Alice's $n$-photon emission 
and Bob's $m$-photon emission separately, 
because  Alice's and Bob's photon sources in the protocols are phase randomized. 
In the following subsections, 
we present the estimation of the phase error rates 
for the cases of Type1 and Type2 independently. 
We derive an upper bound on the phase error $e^{(1,1)}_{i,\U{ph}}$ for $i=1,2$, 
where the superscript $(1,1)$ denotes $n=m=1$ 
and the subscript represents the type of the successful outcome, 
and derive an upper bound on the phase error $e^{(1,2)}_{i,\U{ph}}$. 
We show that in the case of $n=m=2$, no key can be generated 
when the probability of Eve's successful outcome for the two-photon emission event 
is smaller than 1/16. 
We note that in the cases of either $n\geq 3$ or $m\geq 3$, 
Eve can perform an unambiguous state discrimination 
to one of the three-photon emission part~\cite{USD1,USD2}, 
and thus we cannot extract the key from such events, 
given that the channel is lossy enough. 

Fnally, we note that given the phase error rates, 
$Q_i^\U{tot}=\sum_{n,m}Q_i^{(n,m)}$ 
and $e_i^\U{tot}=\sum_{n,m}Q_i^{(n,m)}e^{(n,m)}_{i,\U{bit}}/Q_i^\U{tot}$, 
the asymptotic key rate for Type $i$ is written by~\cite{GLLP} 
\begin{eqnarray}
G_i&=Q_i^{(1,1)}[1-h(e^{(1,1)}_{i,\U{ph}})]+Q_i^{(1,2)}[1-h(e^{(1,2)}_{i,\U{ph}})]
\nonumber\\
&+Q_i^{(2,1)}[1-h(e^{(2,1)}_{i,\U{ph}})]-f(e_i^\U{tot})Q_i^\U{tot}h(e_i^\U{tot}). 
\label{eq:keyrate}
\end{eqnarray}
Here $h(x)=-x\log_{2}x-(1-x)\log_{2}(1-x)$ is the binary shannon entropy. 

\subsubsection{phase error estimation for $(n,m)=(1,1)$ and $(1,2)$}

By the analysis based on the virtual protocol, 
we give the phase error estimation formula for 
$(n,m)=(1,1)$ and $(n,m)=(1,2)$. 
The estimation is performed for Type1 and Type2, separately, 
and we detail the derivation of the phase error estimation in Methods section. 

In the case of Type 1, we have 
\begin{eqnarray}
e^{(1,1)}_{1,\U{ph}}=\frac{3}{2}e^{(1,1)}_{1,\U{bit}} 
\label{eq:11phasebittype1}
\end{eqnarray}
for $(n,m)=(1,1)$ 
and 
\begin{eqnarray}
e^{(1,2)}_{1,\U{ph}}=\min_{s_1}\{s_1\hspace{2pt}e^{(1,2)}_{1,\U{bit}}+f(s_1)\}
\label{eq:12phasebittype1}
\end{eqnarray}
for $(n,m)=(1,2)$, where 
\begin{eqnarray}
f(s_1)=\frac{3-2s_1+\sqrt{6-6\sqrt{2}s_1+4s_1^{2}}}{6}. 
\end{eqnarray}

In the case of Type 2, we have 
\begin{eqnarray}
e^{(1,1)}_{2,\U{ph}}\leq 3e^{(1,1)}_{2,\U{bit}} 
\label{eq:11phasebittype2}
\end{eqnarray}
for $(n,m)=(1,1)$ 
and 
\begin{eqnarray}
e^{(1,2)}_{2,\U{ph}}=\min_{s_2}\{s_2\hspace{2pt}e^{(1,2)}_{2,\U{bit}}+g(s_2)\} 
\label{eq:12phasebittype2}
\end{eqnarray}
for $(n,m)=(1,2)$, where 
$g(s_2)$ is the maximal solution of the following equation for $x$ 
\begin{eqnarray}
&&4\sqrt{2}x^3
+2(1 - 3\sqrt{2} + 3\sqrt{2}s_2) x^2 
\nonumber\\
&+&2(-1 + \sqrt{2} + (1-3\sqrt{2})s_2 + \sqrt{2}s_2^2) x
\nonumber\\
&+&(\sqrt{2}-1)s_2+(1-\sqrt{2})s_2^2
=0. 
\label{eq:gs2}
\end{eqnarray}
\begin{figure}[t] 
\begin{center}
 \includegraphics[width=7cm,clip]{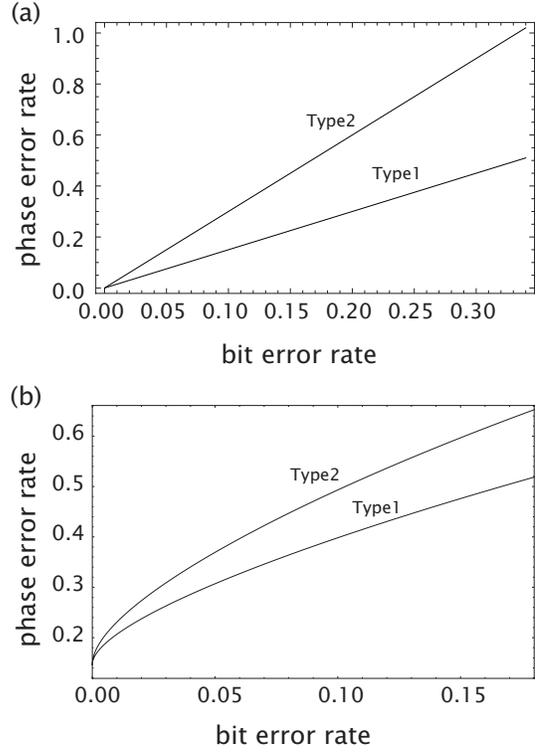}
\end{center}
\caption{
The relations between 
the phase error rates and the bit error rates 
(a) for $(n,m)=(1,1)$ and (b) for $(n,m)=(1,2)$. 
}
\label{fig:pb}
\end{figure}
We depict the dependencies of the phase error rates 
on the bit error rates in Fig.~\ref{fig:pb}. 

\subsubsection{Impossibility of generating a key from $n=m=2$}

For the case of $n=m=2$, 
the key cannot be obtained for $n=m=2$ in Type1 and Type2 
by giving an explicit Eve's attack which give a phase error of 0.5, 
as long as the success probability of Eve's measurement 
conditioned that both Alice and Bob emit two photons is not larger than 1/16. 
We show the proof in Methods section. 
We will prove that we cannot generate a key from $n=m=2$ in the virtual protocol, 
and it follows that we cannot generate a key from $n=m=2$ 
in the actual protocol either. 
To see this, note that the virtual protocol differs from the actual protocol 
only in the way to prepare the state, 
and the state prepared and post data-processing are exactly the same in both protocols. 
In other words, only the local operation needed in state-preparation process 
by the legitimated parties are different in the two protocols. 
By recalling that any local operation cannot convert a separable state 
into a non-separable state, 
we conclude that if we cannot generate a key from a virtual protocol, 
then we cannot generate a key from the actual protocol. 

\subsection{Simulation}
\label{sec:simulation}
Here we show the results of the key generation rate 
for the two experimental setups 
as shown in Figs.~\ref{fig:qnd+parame}(a) and (b)
by using typical experimental parameters 
taken from Gobby-Yuan-Shields~(GYS) experiment~\cite{GYS}, 
where the quantum efficiency and the dark counting 
of the all detectors in Eve's MU are $\eta=0.045$ and $d=8.5\times 10^{-7}$, respectively, 
the loss coefficient of the quantum channel is $\xi = 0.21$dB/km, 
and the inefficiency of the error correcting code is $1.22$. 
In the simulation, 
we use infinite number of decoy states~\cite{decoy} in order to obtain 
$Q_i^{(1,1)}$, $e^{(1,1)}_{i,\U{bit}}$, $Q_i^{(1,2)}$ and $e^{(1,2)}_{i,\U{bit}}$. 
Assuming that the bit error is stemmed only from dark countings of the detectors, 
we ignore the other imperfections such as the misalignment of the devices. 
We also assume that the mean photon numbers of the signal pulses 
prepared by Alice and Bob are the same, 
and the MU in Eve is the middle of Alice and Bob. 
The mean photon number for the signal is optimized 
for maximizing the key generation rate at each distance. 
By using equation~(\ref{eq:keyrate}) 
with the above parameters and assumptions, 
we calculate the key generation rate 
as a function of the distance between Alice and Bob 
(i) 
when Eve postselects the events with $n\leq 1$ and $m\leq 1$ 
with the QND measurement as shown in Fig.~\ref{fig:qnd+parame}(a) 
and Alice and Bob use the coherent pulses, 
and 
(ii) 
when Eve uses the MU in Fig.~\ref{fig:actual} and 
Alice and Bob use quasi single photon sources prepared by the SPDC 
in Fig.~\ref{fig:qnd+parame}(b). 

\begin{figure}[t] 
\begin{center}
 \includegraphics[width=8cm,clip]{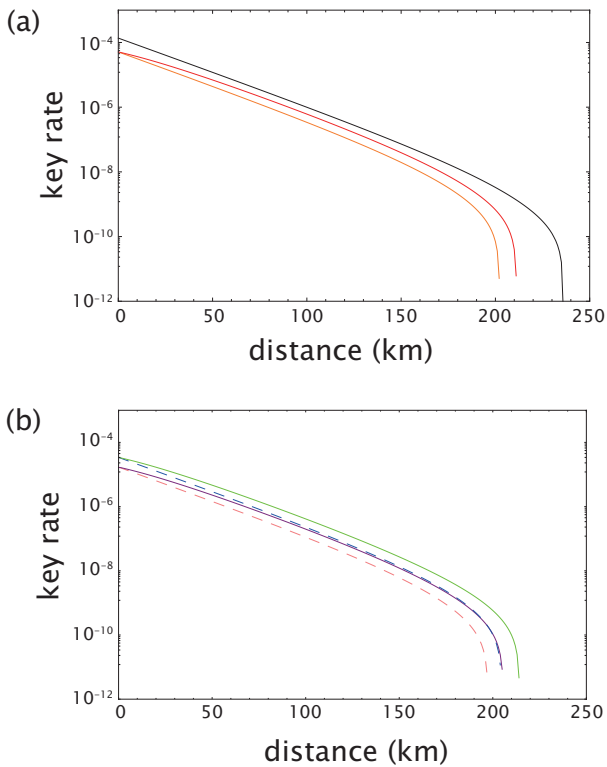}
\end{center}
\caption{
The key rate when Alice and Bob use coherent pulses and 
Eve performs non-destructively exclusion of the multi-photons from Alice and Bob. 
(a) 
Bottom: the key rate of the MDI-SARG04 protocol from $(n,m)=(1,1)$ only. 
Middle: the key rate of the MDI-SARG04 protocol from $(n,m)=(1,1)$, $(1,2)$ and $(2,1)$. 
Top: the key rate of the MDI-BB84 protocol. 
(b)
The upper and lower solid lines are 
the key rates from $(n,m)=(1,1)$, $(1,2)$ and $(2,1)$ for Type1 and Type2, respectively. 
The upper and lower dashed lines are 
the key rates from $(n,m)=(1,1)$ for Type1 and Type2, respectively. 
}
\label{fig:qnd}
\end{figure}
\begin{figure}[t] 
\begin{center}
 \includegraphics[width=7cm,clip]{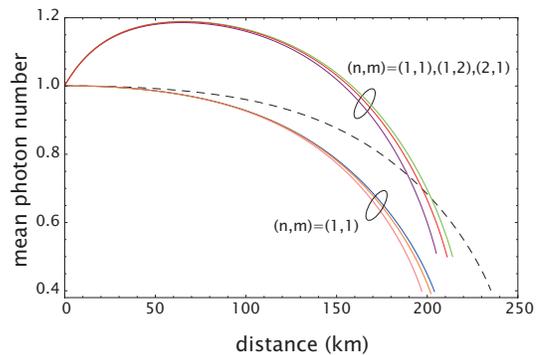}
\end{center}
\caption{
The optimal mean photon number for the key rate in Fig.~\ref{fig:qnd}. 
For the key rates from $(n,m)=(1,1)$, 
the three lines show the mean photon number 
when we consider only Type1, both types and only Type2 from the top. 
The mean photon numbers for the key rates from $(n,m)=(1,1)$, $(1,2)$ and $(2,1)$ 
show a similar tendency. 
The dashed line is for the MDI-BB84 protocol. 
}
\label{fig:qnd_alpha}
\end{figure}

{\it Case (i) -- } 
The simulation result of the key rate is shown in Fig.~\ref{fig:qnd}(a), 
and the mean photon number which maximizes the key rate 
is shown in Fig.~\ref{fig:qnd_alpha}. 
We also plot the key rates of Type1 and Type2 separately 
in Fig.~\ref{fig:qnd}(b). 
The details for obtaining these figures are shown 
in Supplementary. 
When the distance is zero, 
since there is no photon loss before the BS 
and the multi-photon emissions are excluded, 
the events of multi-photon input have no contribution to the key rate. 
In fact, in Fig.~\ref{fig:qnd}(a), 
the two key rates at zero distance obtained from only $(n,m)=(1,1)$ 
and from both $(n,m)=(1,1)$, $(1,2)$ and $(2,1)$ are exactly the same. 
When the distance becomes longer, we see from Fig.~\ref{fig:qnd_alpha} that 
the contribution of the multi photons becomes larger. 
For the key rate from only $(n,m)=(1,1)$, 
the mean photon number is monotonically decrease 
because the multi-photon emissions give only adverse effect. 
On the other hand, when we extract the key additionally from the multi photons, 
the mean photon number does not decrease monotonically, 
which shows an advantage in using multi-photon emission. 

\begin{figure}[t] 
\begin{center}
 \includegraphics[width=7cm,clip]{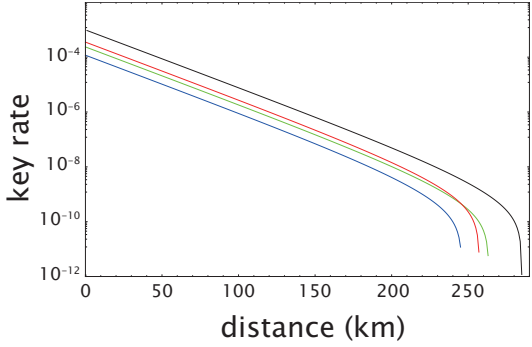}
\end{center}
\caption{
The key rate 
when Alice and Bob use quasi single-photon sources prepared by the SPDC 
and Eve's MU is the same as the circuit used in the MDI-BB84 protocols. 
In this case, the total the key is approximately obtained 
from only the case of $(n,m)=(1,1)$, 
and the successful events of $(n,m)=(1,2)$ and $(n,m)=(2,1)$ 
give little contribution to the key rate. 
This is so because the probability of the two-photon component 
in the heralded photon source 
is negligibly small compared with the probability of the single-photon component. 
The lines are for MDI-BB84~(black), 
for both types~(red), Type1~(green) and Type2~(blue) of the MDI-SARG04. 
}
\label{fig:pdc}
\end{figure}
\begin{figure}[t] 
\begin{center}
 \includegraphics[width=7.5cm,clip]{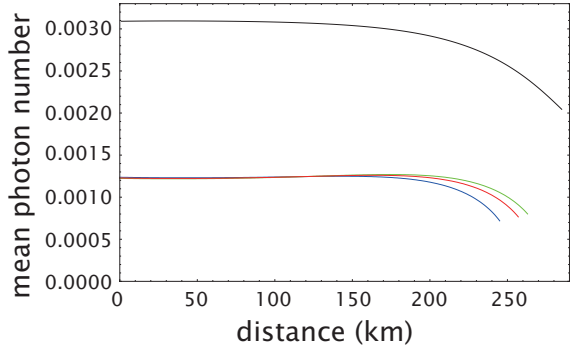}
\end{center}
\caption{
The optimal mean photon number for the key rate in Fig.~\ref{fig:pdc}. 
The upper line~(black) is the mean photon number for the MDI-BB84 protocol. 
The lower three lines are for the key rates of the MDI-SARG04 protocol 
obtained from Type1~(green), both types~(red) and Type2~(blue) from the top. 
}
\label{fig:pdc_alpha}
\end{figure}

{\it Case (ii) -- } 
Alice and Bob use quasi single photon sources by SPDC 
as shown in Fig.~\ref{fig:qnd+parame}(b). 
Detector $\U{D}_0$ is the same as that used in Eve's MU, 
namely it is the threshold detector with 
the quantum efficiency of $\eta=0.045$ 
and the dark counting of $d=8.5\times 10^{-7}$. 
Eve's MU is the same as that shown in Fig.~\ref{fig:actual}. 
The key rate is shown in Fig.~\ref{fig:pdc}. 
The details for calculating the key rates are shown 
in Supplementary. 
The mean photon number which maximizes the key rate 
is shown in Fig.~\ref{fig:pdc_alpha}. 
From Fig.~\ref{fig:pdc}, we see 
that the key rate only from Type1 and that both from Type1 and Type2 intersect. 
For the distribution distance longer than the cross point, 
Type2 has no contribution of the key, 
which is shown by the blue line in the figure, 
and therefore it is better to generate a key from Type1 only. 
From Fig.~\ref{fig:pdc_alpha}, 
we see that the mean photon number is very small. 
This is so because the use of larger mean photon numbers results 
in two-photon emission, which increases the bit error rate. 
From all the figures of the key rate, 
one sees that the key rates of MDI-SARG04 are lower than those of MDI-BB84. 
This tendency holds also for prepare \& measure SARG04~\cite{SARGtamaki,fung}, 
and the higher phase error rates of SARG04 protocol 
than that of BB84 is the main reason of this tendency. 

\section{Discussion}
\label{sec:summary}
In recent years, many proposals and experimental demonstrations 
of the MDI QKD have been studied. 
So far, all of them except for the continuous-variable QKD protocol~\cite{cv} 
are based on the BB84 protocol. 
We first proved the unconditional security of the MDI QKD 
based on the SARG04 protocol. 
In our security proof, 
we gave the upper bounds on the phase error rate 
when Alice and Bob emit single photons and 
when one party emit one photon and the other half emit two photons. 
For the case of the two photon emissions from both parties, 
we proved that a key cannot be generated 
as long as the probability of success in her measurement 
conditioned that both Alice and Bob emit two photons is not larger than 1/16. 
Another important issue to be addressed in MDI setting is 
what kind of measurement should be implemented experimentally 
at Eve's laboratory. 
We have shown that the measurement setup for BB84 in MDI setting 
cannot be used in SARG04 in MDI setting, 
and we proposed two measurement schemes for MDI SARG04. 
In the first one, 
Alice and Bob use heralded single photon sources prepared by SPDC. 
In the second one, 
Eve performs QND measurement on the two pulses 
coming from Alice and Bob individually. 
In our simulation based on these experimental setups, 
it was confirmed that these setup can generate a key. 

\section{Methods}

\subsubsection{Proof of phase error estimation for $n=m=1$}
\label{11}
\begin{figure}[t] 
\begin{center}
 \includegraphics[width=8.5cm,clip]{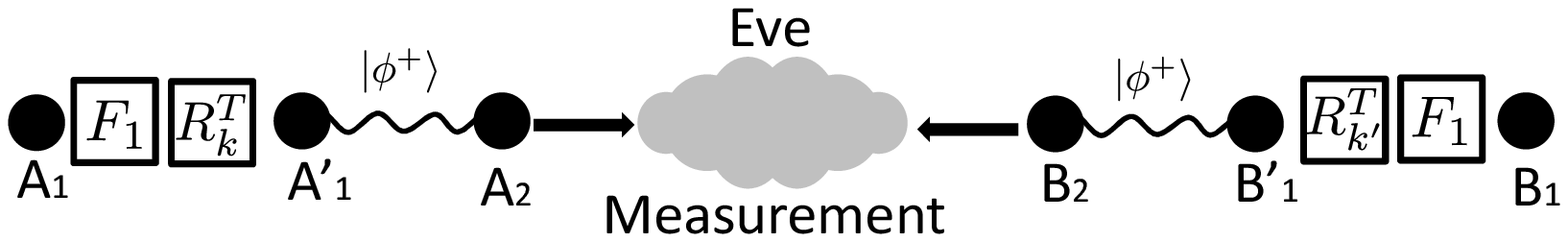}
\end{center}
\caption{
Schematic that is equivalent to the EDP for $n=m=1$. 
While 
Eve accesses only the photons in modes $\U{A}_2$ and $\U{B}_2$ 
in the actual protocol, 
we pessimistically suppose that she can prepare any state in $\U{A}_1'$ and $\U{B}_1'$ 
for simplicity of the proof. 
}
\label{fig:nm11}
\end{figure}
Here, we give the phase error estimation for $n=m=1$. 
For this, it is convenient to recall a mathematical property 
of the maximally entangled state that 
$(I\u{1}\otimes M\u{2})\ket{\phi^+}\u{12}
=(M\u{1}^T\otimes I\u{2})\ket{\phi^+}\u{12}$ is satisfied 
for any operator $M$. 
Therefore $\ket{\Phi_{1,k}}\u{A_1A_2}$ in (v1) is expressed as 
$\ket{\Phi_{1,k}}\u{A_1A_2}\propto 
F_{1,\U{A'_1}}R^T_{k,\U{A'_1}}\ket{\phi^+}\u{A'_1A_2}$, 
where 
$F_{1,\U{A'_1}}=\cos(\pi/8)\ket{0_{x}}\u{A_1}\bra{{0_{x}}}\u{A_1'}+
\sin(\pi/8)\ket{1_{x}}\u{A_1}\bra{{1_{x}}}\u{A_1'}$. 
Physically, this identification can be interpreted as the situation 
where $\ket{\phi^{+}}$ is prepared by each of the parties, 
the filtering operation, 
of which successful case is described by $F_1$, is applied, 
and then each party sends the photons to Eve 
only when the filtering operation succeeds~(also see Fig.~\ref{fig:nm11}). 
For the simplicity of the security proof, 
we make an overestimation of Eve's ability in terms of the accessibility 
of the photons, namely, we imagine Eve who has a direct access 
to photons of $\U{A}_1'$ and $\U{B}_1'$ rather than $\U{A}_2$ and $\U{B}_2$, 
and she can prepare any joint state of the photons of $\U{A}_1'$ and $\U{B}_1'$. 
For later convenience, 
we denote by $\rho^{(1,1)}\u{A'_1B'_1|suc}$ the state prepared by Eve. 

In the following, we first discuss the case of Type1. 
We define $\tilde{e}^{(1,1)}_{1,\U{bit/ph}}
=\tr(\Pi^{(1,1)}_{1,\U{bit/ph}}\rho^{(1,1)}\u{A'_1B'_1|suc})$ 
as the joint probability that the photons in $\rho^{(1,1)}\u{A'_1B'_1|suc}$ 
pass through the filtering operation and induces a bit/phase error 
to the state $\ket{\psi^-}$ after the rotation.
Here $\Pi^{(1,1)}_{1,\U{bit}}$ and $\Pi^{(1,1)}_{1,\U{ph}}$ are 
POVM elements of the bit and phase error measurements on $\rho^{(1,1)}\u{A'_1B'_1|suc}$, 
respectively. 
The probability that the two photons in $\rho^{(1,1)}\u{A'_1B'_1|suc}$ pass 
through the successful filtering operation is described by 
$p^{(1,1)}_{1,\U{fil}}=\tr(\Pi^{(1,1)}_{1,\U{fil}}\rho^{(1,1)}\u{A'_1B'_1|suc})$, 
where the POVM element of the successful filtering operation on the two photons is 
\begin{eqnarray}
{\Pi}^{(1,1)}_{1,\U{fil}}=\frac{1}{4}
\sum_{k=0}^{3}
{P}(R_{k,\U{A'_1}}F^T_{1,\U{A'_1}}
R_{k,\U{B'_1}}F^T_{1,\U{B'_1}}), 
\label{eq:pifil11type1} 
\end{eqnarray}
where ${P}(\cdot) = {(\cdot)(\cdot)^{\dag}}$. 
The POVMs for the bit and the phase errors are written as 
\begin{eqnarray}
\Pi^{(1,1)}_{1,\U{bit/ph}}&=&\frac{1}{4}
\sum_{i=0}^{1}\sum_{k=0}^{3}
P(R_{k,\U{A'_1}}F^T_{1,\U{A'_1}}\ket{i_{z/x}}\u{A_1}\nonumber\\
&&\otimes R_{k,\U{B'_1}}F^T_{1,\U{B'_1}}\ket{i_{z/x}}\u{B_1}). 
\label{eq:pibit11type1}
\end{eqnarray}
Applying the Bayes' rule, 
the bit error rate $e^{(1,1)}_{1,\U{bit}}$ and 
the phase error rate $e^{(1,1)}_{1,\U{ph}}$ in the final state 
in modes $\U{A_1}$ and $\U{B_1}$ are described by 
\begin{eqnarray}
e^{(1,1)}_{1,\U{bit/ph}}
=\frac{\tilde{e}^{(1,1)}_{1,\U{bit/ph}}}{p^{(1,1)}_{1,\U{fil}}}. 
\label{eq:phasebitperfil11type1}
\end{eqnarray}
The phase error estimation can be established by directly writing down 
the explicit form of equation~(\ref{eq:pibit11type1}) 
comparing each matrix element, and one can conclude that 
\begin{eqnarray}
{\Pi}^{(1,1)}_{1,\U{ph}}=\frac{3}{2}{\Pi}^{(1,1)}_{1,\U{bit}}. 
\end{eqnarray} 
Thus from equations~(\ref{eq:pifil11type1}) and (\ref{eq:phasebitperfil11type1}), 
the phase error rate is precisely estimated, 
by using the bit error rate, as shown in equation~(\ref{eq:11phasebittype1}). 
Thanks to Azuma's inequality~\cite{Azuma}, 
equation~(\ref{eq:11phasebittype1}) holds for any eavesdropping including 
coherent attacks.

Next, we estimate the phase error rate for Type2. 
Because only the cases of $k=k'=0,2$ are accepted for Type2, 
the definition of the POVM element of the successful filtering operation 
is changed to 
\begin{eqnarray}
{\Pi}^{(1,1)}_{2,\U{fil}}=\frac{1}{2}
\sum_{k=0,2}
P(R_{k,\U{A_1'}}F^T_{1,\U{A_1'}}
R_{k,\U{B_1'}}F^{T}_{1,\U{B_1'}}), 
\label{eq:pifil11type2}
\end{eqnarray}
and the probability that the two photons in $\rho^{(1,1)}\u{A'_1B'_1|suc}$ pass 
through the successful filtering operation is expressed by 
$p^{(1,1)}_{2,\U{fil}}=\tr(\Pi^{(1,1)}_{2,\U{fil}}\rho^{(1,1)}\u{A'_1B'_1|suc})$. 
We describe a joint probability that 
the two photons in $\rho^{(1,1)}\u{A'_1B'_1|suc}$ pass 
through the successful filtering operation after the rotation and 
then the photons in modes $\U{A_1}$ and $\U{B_1}$ have a bit/phase error 
to the state $\ket{\psi^+}$ 
by 
$\tilde{e}^{(1,1)}_{2,\U{bit/ph}}
=\tr(\Pi^{(1,1)}_{2,\U{bit/ph}}\rho^{(1,1)}\u{A'_1B'_1|suc})$. 
Like in the case of Type1, 
the POVM elements of $\Pi^{(1,1)}_{2,\U{bit}}$ and $\Pi^{(1,1)}_{2,\U{ph}}$ 
are written by 
\begin{eqnarray}
\Pi^{(1,1)}_{2,\U{bit}}&=&\frac{1}{2}
\sum_{i=0}^{1}\sum_{k=0,2}
P(R_{k,\U{A'_1}}F^T_{1,\U{A'_1}}\ket{i_{z}}\u{A_1} \nonumber\\
&&\otimes R_{k,\U{B'_1}}F^T_{1,\U{B'_1}}\ket{{i\oplus 1}_{z}}\u{B_1})
\label{eq:pibit11type2}
\end{eqnarray}
and 
\begin{eqnarray}
\Pi^{(1,1)}_{2,\U{ph}}&=&\frac{1}{2}
\sum_{i=0}^{1}\sum_{k=0,2}
P(R_{k,\U{A'_1}}F^T_{1,\U{A'_1}}\ket{i_{x}}\u{A_1} \nonumber\\
&&\otimes R_{k,\U{B'_1}}F^T_{1,\U{B'_1}}\ket{i_{x}}\u{B_1}). 
\label{eq:piph11type2}
\end{eqnarray}

By using the Bayes' rule, 
the bit/phase error rate of $e^{(1,1)}_{2,\U{bit/ph}}$ 
in the final state is expressed by 
\begin{eqnarray}
e^{(1,1)}_{2,\U{bit/ph}}
=\frac{\tilde{e}^{(1,1)}_{2,\U{bit/ph}}}{p^{(1,1)}_{2,\U{fil}}}. 
\label{eq:phasebitperfil11type2}
\end{eqnarray}
In order to see the relation between the bit and phase error rates, 
we consider an inequality to bound the phase error as 
$se^{(1,1)}_{2,\U{bit}}-e^{(1,1)}_{2,\U{ph}}\geq 0$, 
where $s$ is a real number, 
which is equivalent to 
$s\Pi^{(1,1)}_{2,\U{bit}}-\Pi^{(1,1)}_{2,\U{ph}}\geq 0$ 
for $p^{(1,1)}_{2,\U{fil}}>0$. 
By considering a non-negativity condition of 
$s\Pi^{(1,1)}_{2,\U{bit}}-\Pi^{(1,1)}_{2,\U{ph}}\geq 0$,  
we see that this inequality always holds when $s\geq 3$, 
and therefore, we have the relation 
between the phase error rate and the bit error 
as shown in equation~(\ref{eq:11phasebittype2}). 

\subsubsection{Proof of phase error estimation for $n=1$ and $m=2$}
\label{12}
\begin{figure}[t] 
\begin{center}
 \includegraphics[width=8.5cm,clip]{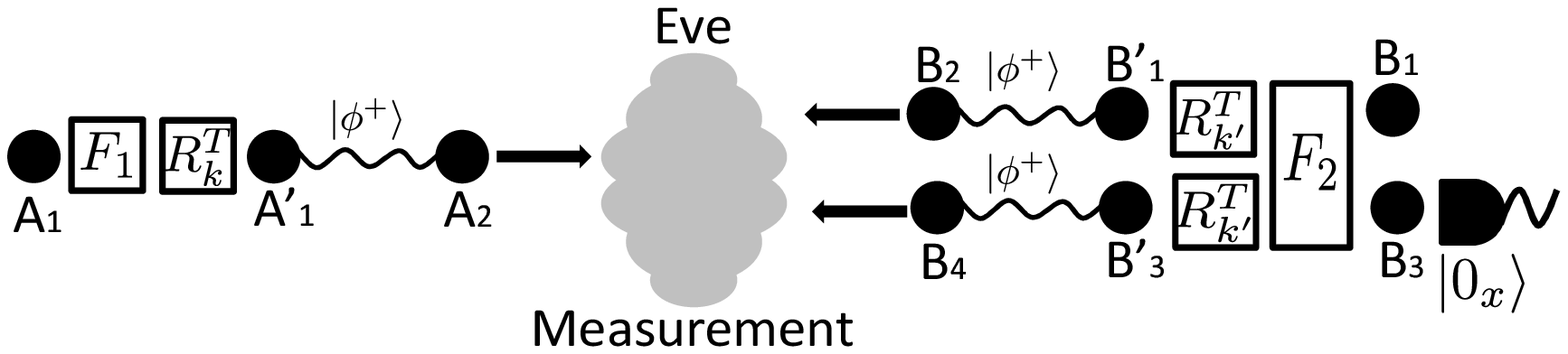}
\end{center}
\caption{
Schematic which is equivalent to the EDP for $n=1$ and $m=2$. 
By Eve's announcement 
for the successful measurement on the photons in 
$\U{A_2}$, $\U{B_2}$ and $\U{B_4}$, 
the three-photon state 
$\rho^{(1,2)}\u{A'_1B'_1B'_3|suc}$ is prepared. 
}
\label{fig:nm12}
\end{figure}

Below, we give the phase error estimation for $n=1$ and $m=2$. 
By using the similar argument as $n=m=1$, 
$\ket{\Phi_{2,k}}\u{B_1, B_2}$ at Bob's side in (v1) is defined by 
$\bra{0_x}\u{B_3}F_{2,\U{B'_1B'_3}}R^T_{k,\U{B'_1}}R^T_{k,\U{B'_3}}
\ket{\phi^+}\u{B'_1B_2}\ket{\phi^+}\u{B'_3B_4}$ as in Fig.~\ref{fig:nm12}, 
where 
$F_{2,\U{B'_1B'_3}}=\cos^2(\pi/8)\ket{0_x0_x}\u{B_1B_3}\bra{0_x0_x}\u{B'_1B'_3}
+\sin^2(\pi/8)\ket{0_x0_x}\u{B_1B_3}\bra{1_x1_x}\u{B'_1B'_3}
+\sqrt{2}\cos(\pi/8)\sin(\pi/8)\ket{1_x0_x}\u{B_1B_3}\bra{\psi^+}\u{B'_1B'_3}$. 
Here we note that two-photon emission part is simulated 
by preparing two pairs of $\ket{\phi^{+}}$ followed by 
the rotation and the filtering operation 
on two qubits~(see also Fig.~\ref{fig:nm12}). 
In this virtual protocol, while 
we consider two photons in different modes, 
this never underestimates Eve's ability. 
This is so because two photons in the different modes 
and two photons in a single mode can be converted 
just by an unitary transformation as 
$\ket{\varphi_i}\u{B_2}\ket{\varphi_i}\u{B_4}
\rightarrow \ket{2_{\varphi_i}}\u{B_2}$. 
We note that 
because the photon in mode $\U{B}_3$ is in $\ket{0_x}$ after the filtering operation, 
and it is decoupled from all the other systems, 
the component is not related to the security proof. 
Again, we employ the overestimation that Eve has the control over the state 
of the systems of $\U{A'_1}$, $\U{B'_1}$ and $\U{B'_3}$, 
and we denote the three-photon state by $\rho^{(1,2)}\u{A'_1B'_1B'_3|suc}$, 
which is prepared by Eve after her announcement of the success.
Like in the case for $n=m=1$, 
we estimate a phase error for each case of Type1 and Type2 separately. 

For Type1, 
define a POVM element of the successful filtering operations 
on $\rho^{(1,2)}\u{A'_1B'_1B'_3|suc}$ as 
\begin{eqnarray}
{\Pi}^{(1,2)}_{1,\U{fil}}=\frac{1}{4}\sum_{k=0}^{3}
P(R_{k,\U{A'_1}}F^T_{1,\U{A'_1}} R_{k,\U{B'_1}}R_{k,\U{B'_3}}F_{2,\U{B'_1B'_3}}^T). 
\label{eq:pifil12type1}
\end{eqnarray}
Here the probability of the successful filtering operation is written by 
$p^{(1,2)}_{1,\U{fil}}=\tr(\Pi^{(1,2)}_{1,\U{fil}}\rho^{(1,2)}\u{A'_1B'_1B'_3|suc})$. 
We define 
$\tilde{e}^{(1,2)}_{1,\U{bit/ph}} 
=\tr(\Pi^{(1,2)}_{1,\U{bit/ph}}\rho^{(1,2)}\u{A'_1B'_1B'_3|suc})$ 
as a joint probability that 
the photons in $\rho^{(1,2)}\u{A'_1B'_1B'_3|suc}$ 
pass through the filtering operation and induces a bit/phase error 
to the state $\ket{\psi^-}$ after the rotation.
the successful filtering operation after the rotation is performed 
on the two photons in $\rho^{(1,2)}\u{A'_1B'_1B'_3|suc}$ and 
then the photons in modes $\U{A_1}$ and $\U{B_1}$ have a bit/phase error 
to the state $\ket{\psi^-}$. 
Here, POVM elements of ${\Pi}^{(1,2)}_{1,\U{bit/ph}}$ are written by 
\begin{eqnarray}
{\Pi}^{(1,2)}_{1,\U{bit/ph}}&=&\frac{1}{4}\sum_{i=0}^{1}\sum_{k=0}^{3}
P(R_{k,\U{A'_1}}F^T_{1,\U{A'_1}}\ket{i_{z/x}}\u{A_1} \nonumber\\
&&\otimes
R_{k,\U{B'_1}}R_{k,\U{B'_3}}F_{2,\U{B'_1B'_3}}^T\ket{i_{z/x}}\u{B_1}\ket{0_x}\u{B_3}). 
\label{eq:pibit12type1}
\end{eqnarray}
The actual bit error rate $e^{(1,2)}_{1,\U{bit}}$ 
and phase error rate $e^{(1,2)}_{1,\U{ph}}$ 
for $n=1$ and $m=2$ are obtained by accommodating the normalization 
by $p\u{1,fil}^{(1,2)}$, and they are expressed as
\begin{eqnarray}
e^{(1,2)}_{1,\U{bit/ph}}=\frac{\tilde{e}^{(1,2)}_{1,\U{bit/ph}}}{p^{(1,2)}_{1,\U{fil}}}. 
\label{eq:eebit12type1}
\end{eqnarray}
In order to see the relation between the bit and phase error rates,
we consider an inequality to bound the phase error as 
$e^{(1,2)}_{1,\U{ph}}\leq s_{1}e^{(1,2)}_{1,\U{bit}}+t_{1}$, 
where $s_{1}$ and $t_{1}$ are real numbers. 
By using equations~(\ref{eq:pifil12type1}) -- (\ref{eq:eebit12type1}), 
and the linearity of the trace, we obtain an inequality as 
$s_{1}{\Pi}^{(1,2)}_{1,\U{bit}}
+t_{1}{\Pi}^{(1,2)}_{1,\U{fil}}-{\Pi}^{(1,2)}_{1,\U{ph}}\geq{0}$, 
which is satisfied when 
\begin{eqnarray}
t_1\geq{f(s_1)}=\frac{3-2s_1+\sqrt{6-6\sqrt{2}s_1+4s_1^{2}}}{6}. 
\end{eqnarray}
Therefore the phase error rate is given by using the bit error as 
shown in equation~(\ref{eq:12phasebittype1}). 

For Type2, we define a POVM element ${\Pi}^{(1,2)}_{2,\U{fil}}$ 
of the successful filtering operation 
by limiting the summation only to $k=0, 2$ 
and by replacing $1/4$ with $1/2$ in equation~(\ref{eq:pifil12type1}). 
The probability of the successful filtering operation is described by 
$p_{2,\U{fil}}^{(1,2)}=\tr({\Pi}^{(1,2)}_{2,\U{fil}}\rho^{(1,2)}\u{A'_1B'_1B'_3|suc})$. 
We also define joint probabilities of 
$\rho^{(1,2)}\u{A'_1B'_1B'_3|suc}$ passing through the filtering and 
presenting bit and phase errors to the state $\ket{\psi^+}$ by 
$\tilde{e}^{(1,2)}_{2,\U{bit}} 
=\tr(\Pi^{(1,2)}_{2,\U{bit}}\rho^{(1,2)}\u{A'_1B'_1B'_3|suc})$ 
and 
$\tilde{e}^{(1,2)}_{2,\U{ph}} 
=\tr(\Pi^{(1,2)}_{2,\U{ph}}\rho^{(1,2)}\u{A'_1B'_1B'_3|suc})$. 
We define the POVM element of ${\Pi}^{(1,2)}_{2,\U{ph}}$ by 
limiting the summation only to $k=0, 2$ and replacing $1/4$ with $1/2$ 
in equation~(\ref{eq:pibit12type1}), 
and that of ${\Pi}^{(1,2)}_{2,\U{bit}}$ is defined by 
limiting the summation only to $k=0, 2$, 
replacing $1/4$ with $1/2$, and 
$\ket{i_z}$ with $\ket{i\oplus 1_z}$ for mode $\U{B_1}$. 
In a similar manner as the case of Type1 for $n=1$ and $m=2$, 
by using 
the bit error rate defined by 
$e^{(1,2)}_{2,\U{bit}}=\tilde{e}^{(1,2)}_{2,\U{bit}}/p^{(1,2)}_{2,\U{fil}}$, 
the phase error rate as 
$e^{(1,2)}_{2,\U{ph}}=\tilde{e}^{(1,2)}_{2,\U{ph}}/p^{(1,2)}_{2,\U{fil}}$ 
and real numbers $s_{2}$ and $t_{2}$, 
we consider an inequality as 
$e^{(1,2)}_{2,\U{ph}}\leq s_{2}e^{(1,2)}_{2,\U{bit}}+t_{2}$, 
which leads to 
$s_{2}{\Pi}^{(1,2)}_{2,\U{bit}}
+t_{2}{\Pi}^{(1,2)}_{2,\U{fil}}-{\Pi}^{(1,2)}_{2,\U{ph}}\geq{0}$. 
From this inequality, we obtain $t_2 \geq g(s_2)$, 
where $g(s_2)$ is the maximal solution of equation~(\ref{eq:gs2}). 
Using $g(s_2)$, we have the relation between the phase error rate and 
the bit error as shown in equation~(\ref{eq:12phasebittype2}). 

\subsubsection{Proof of impossibility of generating a key from $n=m=2$}
\label{22}
\begin{figure}[t] 
\begin{center}
 \includegraphics[width=8.5cm,clip]{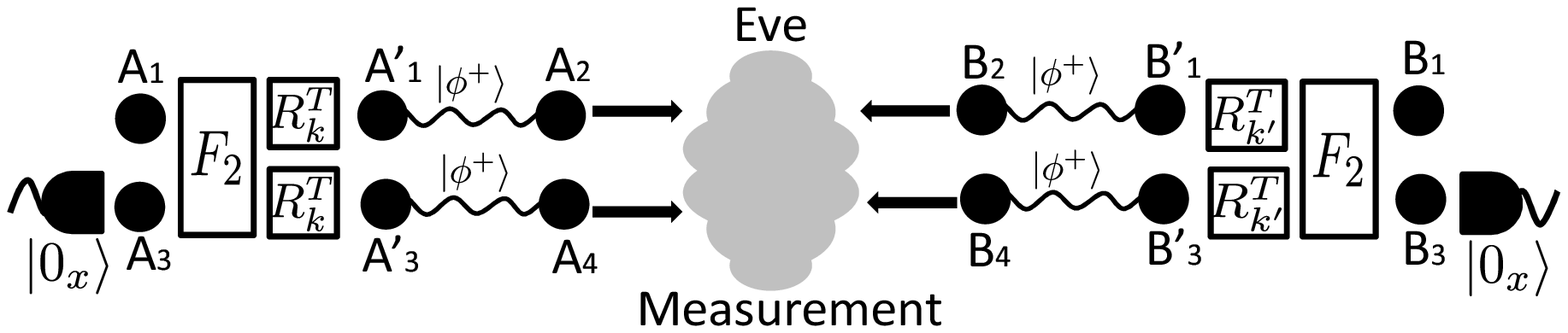}
\end{center}
\caption{
Schematic which is equivalent to the EDP for $n=m=2$. 
By Eve's announcement 
for the successful measurement on the photons in 
$\U{A_2}$, $\U{A_4}$, $\U{B_2}$ and $\U{B_4}$, 
the four-photon state 
$\rho^{(2,2)}\u{A'_1A'_3B'_1B'_3|suc}$ is prepared. 
}
\label{fig:nm22}
\end{figure}

For the case of $n=m=2$, 
like in the previous subsection, 
$\ket{\Phi_{2,k}}\u{A_1, A_2}$ at Alice's side in (v1) is obtained by 
$\bra{0_x}\u{A_3}F_{2,\U{A'_1A'_3}}R^T_{k,\U{A'_1}}R^T_{k,\U{A'_3}}
\ket{\phi^+}\u{A'_1A_2}\ket{\phi^+}\u{A'_3A_4}$, 
and $\ket{\Phi_{2,k}}\u{B_1, B_2}$ at Bob's side is prepared by the same manner. 
As a result, the virtual protocol for $n=m=2$ is equivalent 
to the successful situation of the filtering operations, 
which we depict in Fig.~\ref{fig:nm22}. 
We denote the state of Alice's and Bob's four qubits 
after Eve's successful announcement by $\rho^{(2,2)}\u{A_1'A_3'B_1'B_3'|suc}$. 
In the following, we prove that the key cannot be obtained 
for $n=m=2$ by giving an explicit Eve's attack, namely 
we give explicit states of $\U{A'_1}$, $\U{A'_3}$, $\U{B'_1}$ and $\U{B'_3}$ 
which give a phase error of 0.5. 
The key ingredient is that while Eve cannot manipulate these four qubits, 
she conclusively prepare such a state on their qubits 
by announcing the success of her measurement only 
when she succeeds an eavesdropping measurement 
on Eve's photons $\U{A_2}$, $\U{A_4}$, $\U{B_2}$ and $\U{B_4}$. 
This attack gives Eve the perfect information on the bit values 
when her measurement succeeds. 

For Type1, the probability of the successful filtering operation is 
expressed by 
$p^{(2,2)}_{1,\U{fil}}=\tr(\Pi^{(2,2)}_{1,\U{fil}}\rho^{(2,2)}\u{A_1'A_3'B_1'B_3'|suc})$,
where 
\begin{align}
{\Pi}^{(2,2)}_{1,\U{fil}}=\frac{1}{4}
\sum_{k=0}^{3}
P(R_{k,\U{A'_1}}R_{k,\U{A'_3}}F^{T}_{2,\U{A'_1A'_3}}
R_{k,\U{B'_1}}R_{k,\U{B'_3}}F^{T}_{2,\U{B'_1B'_3}}). \nonumber\\
\label{eq:pifil22}
\end{align}
The joint probability, 
that the filtering operation succeeds 
and the bit/phase error to the state $\ket{\psi^-}$ is detected, 
is expressed by 
$\tilde{e}^{(2,2)}_{1,\U{bit/ph}}
=\tr(\Pi^{(2,2)}_{1,\U{bit/ph}}\rho^{(2,2)}\u{A_1'A_3'B_1'B_3'|suc})$, 
where 
\begin{eqnarray}
{\Pi}^{(2,2)}_{1,\U{bit/ph}}
&=&\frac{1}{4}\sum_{i=0}^{1}\sum_{k=0}^{3}
P(R_{k,\U{A'_1}}R_{k,\U{A'_3}}F^{T}_{2,\U{A'_1A'_3}}
\ket{i_{z/x}}\u{A_1}\ket{0_x}\u{A_3} \nonumber \\
&&\otimes R_{k,\U{B'_1}}R_{k,\U{B'_3}}F^{T}_{2,\U{B'_1B'_3}}
\ket{i_{z/x}}\u{B_1}\ket{0_x}\u{B_3}). 
\label{eq:pibit22}
\end{eqnarray}
The bit/phase error rate is expressed as 
$e^{(2,2)}_{1,\U{bit/ph}}=\tilde{e}^{(2,2)}_{1,\U{bit/ph}}/p^{(2,2)}_{1,\U{fil}}$. 
One can confirm by direct calculation that 
a four-photon state of 
$\ket{\mu_1}\u{A_1'B_1'A_3'B_3'}=\ket{\psi^-}\u{A_1'B_3'}\ket{0_x1_x}\u{A_3'B_1'}$ 
gives 
$e^{(2,2)}_{1,\U{bit}}=0$ and $e^{(2,2)}_{1,\U{ph}}=0.5$, 
and another four-photon state 
$\ket{\mu_2}\u{A_1'B_1'A_3'B_3'}=(\ket{0_z0_z1_z0_z}\u{A_1'A_3'B_1'B_3'}
+\ket{1_z0_z0_z1_z}\u{A_1'A_3'B_1'B_3'})/\sqrt{2}$, 
which is orthogonal to $\ket{\mu_1}\u{A_1'B_1'A_3'B_3'}$, 
gives 
$e^{(2,2)}_{1,\U{bit}}=0.5$ and $e^{(2,2)}_{1,\U{ph}}=0.5$. 
Therefore, although Eve cannot touch the four modes 
$\U{A'_1}$, $\U{B'_1}$, $\U{A'_3}$ and $\U{B'_3}$, 
Eve can prepare the two states 
by a projective measurement on 
the four photons in $\U{A_2}$, $\U{B_2}$, $\U{A_4}$ and $\U{B_4}$ 
as $\{ P(\ket{\mu_1}), P(\ket{\mu_2}), I-\sum_{i=1}^2P(\ket{\mu_i}) \}$. 
One sees this fact from the equation 
${}\u{A_2B_2A_4B_4}\expect{\mu_i|\phi^+}\u{A_1'A_2}
\ket{\phi^+}\u{B_1'B_2}
\ket{\phi^+}\u{A_3'A_4}
\ket{\phi^+}\u{B_3'B_4}
=\ket{\mu_i}\u{A_1'B_1'A_3'B_3'}/\sqrt{16}$, 
which also implies that the preparation succeeds with a probability of 1/16. 
Thus a malicious Eve achieves the phase error rate of 0.5 
for any bit error rate by distributing these states with a relevant probability.
This means that the state in $\U{A}_1$ and $\U{B}_1$ is separable, 
and it follows that no key can be generated for $q_1^{(2,2)}\leq 1/16$, 
where $q_i^{(2,2)}$ is the probability  of 
Eve's successful detection of Type $i$ conditioned that 
both Alice and Bob emit two photons. 

For Type2, with the same fashion as the case of $n=1$ and $m=2$, 
POVM elements ${\Pi}^{(2,2)}_{2,\U{fil}}$ 
for the successful filtering operation and 
${\Pi}^{(2,2)}_{2,\U{bit/ph}}$ for the bit/phase error are defined 
by replacing the summation range of $k$, 
the prefactor and the proper inversion of the bit value of the projection 
in equations~(\ref{eq:pifil22}) and (\ref{eq:pibit22}). 
We consider the following four orthogonal four-photon states 
for systems $\U{A'_1}$, $\U{B'_1}$, $\U{A'_3}$ and $\U{B'_3}$ 
$\ket{\nu_1}\u{A_1'B_1'A_3'B_3'}=\ket{\psi^+}\u{A'_1B'_1}\ket{0_x0_x}\u{A'_3B'_3}$, 
$\ket{\nu_2}\u{A_1'B_1'A_3'B_3'}=\ket{\psi^+}\u{A'_1B'_1}\ket{0_x1_x}\u{A'_3B'_3}$, 
$\ket{\nu_3}\u{A_1'B_1'A_3'B_3'}=\ket{\psi^+}\u{A'_1B'_1}\ket{1_x0_x}\u{A'_3B'_3}$ and 
$\ket{\nu_4}\u{A_1'B_1'A_3'B_3'}=\ket{\psi^-}\u{A'_1B'_1}\ket{0_x0_x}\u{A'_3B'_3}$. 
Each state can be prepared by Eve's projective measurement 
$\{ P(\ket{\nu_1}),P(\ket{\nu_2}),P(\ket{\nu_3}), P(\ket{\nu_4}),
I-\sum_{i=1}^4P(\ket{\nu_i})\}$
on the four photons in $\U{A_2}$, $\U{B_2}$, $\U{A_4}$ and $\U{B_4}$ 
with a probability of $1/16$. 
By calculating the error probabilities, 
we see that mixed states 
$0.25\ketbra{\nu_1}{\nu_1}+0.75\ketbra{\nu_2}{\nu_2}$ 
and $0.75\ketbra{\nu_3}{\nu_3}+0.25\ketbra{\nu_4}{\nu_4}$ give 
$(e^{(2,2)}_{2,\U{bit}}, e^{(2,2)}_{2,\U{ph}})=(0, 0.5)$ and 
$(e^{(2,2)}_{2,\U{bit}}, e^{(2,2)}_{2,\U{ph}})=(0.5, 0.5)$, respectively. 
Therefore Eve achieves any bit error rate below 0.5 
while keeping $e^{(2,2)}_{2,\U{ph}}=0.5$ by distributing 
the above two mixed states with an appropriate probability. 
As a result, we conclude that for $q_2^{(2,2)}\leq 1/16$, 
the key cannot be obtained. 

\section*{Acknowledgements}
This work was supported by the Funding Program 
for World-Leading Innovative R \& D on Science and Technology~(FIRST), 
MEXT Grant-in-Aid for Scientific Research 
on Innovative Areas 21102008, 
MEXT Grant-in-Aid for Young scientists(A) 23684035, 
JSPS Grant-in-Aid for Scientific Research(A) 25247068 and (B) 25286077. 
KT acknowledges support from the National Institute of Information 
and Communications Technology~(NICT) of Japan~(project 
``Secure photonic network technology'' as part of ``The project UQCC'').

\newpage

\begin{widetext}

\section{Supplementary material: Details of the simulation}
\label{sec:app}
Here we describe the details of our simulation. 
The quantum efficiency and the dark counting 
of the detectors are $\eta=0.045$ and $d=8.5\times 10^{-7}$, respectively. 
The loss coefficient of the quantum channel is $\xi = 0.21$dB/km. 
We denote $a=\cos(\pi/8)$ and $b=\sin(\pi/8)$. 
We define that $p^{(n,m)}_{i,ab}$ is a probability that 
the photons are detected as the successful event of Type $i$ 
conditioned that Alice and Bob emit $n$ and $m$ photons in the states 
$\ket{\varphi_a}$ and $\ket{\varphi_b}$ for $a,b=0,\ldots, 3$, 
respectively. 
$q_{i,ab}$ is the probability of the successful detection of Type $i$ 
conditioned that 
Alice and Bob emit photons in $\ket{\varphi_a}$ and $\ket{\varphi_b}$, respectively.
Assuming that Eve is in the middle of Alice and Bob, 
the channel transmittance to Eve from Alice is the same as that from Bob. 
Denoting that $l$ is the distance between Alice and Bob, 
the channel transmittance for Alice and Bob is 
\begin{eqnarray}
T=10^{-\xi{0.5l/10}}. 
\end{eqnarray}
In the following, we give the experimental data for the simulation 
(i) when Eve postselects the events with $n\leq 1$ and $m\leq 1$ 
by the QND measurement before mixing the pulses 
from Alice and Bob~(see Fig.~2(a)), 
and 
(ii) 
when Alice and Bob use quasi single photon sources 
by the SPDC~(see Fig.~2(b)). 

\subsection{Case (i) Eve performs the QND measurement.}
Each of Alice and Bob uses a phase randomized weak coherent pulse 
with the mean photon number of $\mu$. 
The probability $p_{n}$ for $n$-photon emission is 
\begin{eqnarray}
p_{n}(\mu)=e^{-\mu}\frac{\mu^n}{n!}. 
\end{eqnarray}
For later use, we define the equations
\begin{eqnarray}
f_1&=&(1-d)^2(
2\eta^2a^2b^2(1+3d) +2\eta(1-\eta) d +2(1-\eta )^2d^2),\\
f_2&=&f_1-(1-d)^22a^2b^2\eta^2, \\
f_3&=&(1-d)^2(\eta d+2(1-\eta )d^2),\\
f_4&=&(1-d)^22d^2, \\
f_5&=&(1-d)^2(2\eta^2a^2b^2(1+d) +2\eta(1-\eta) d +2(1-\eta )^2d^2).\nonumber\\
\end{eqnarray}
In the following, we give $Q^{(n,m)}_i$ and $e\u{i,bit}^{(n,m)}$. 

For Type1, we have 
\begin{eqnarray}
Q^{(1,1)}_{1}
&=&p_1^2(\mu)(2p^{(1,1)}_{1,00}+p^{(1,1)}_{1,01}+p^{(1,1)}_{1,12})/4,\\
e\u{1,bit}^{(1,1)}&=&p_1^2(\mu)p^{(1,1)}_{1,00}/(2 Q^{(1,1)}_{1}),\\
Q^{(1,2)}_1&=&p_1(\mu)p_{2}(\mu)(2p^{(1,2)}_{1,00}+p^{(1,2)}_{1,01}+p^{(1,2)}_{1,12})/4,
\\
e^{(1,2)}\u{1,bit}&=&p_1(\mu)p_{2}(\mu)p^{(1,2)}_{1,00}/(2Q^{(1,2)}_{1}),\\
Q^\U{tot}_{1}&=&
(2q_{1,00}+q_{1,01}+q_{1,12})/4,\\
e^\U{tot}_{1}&=&q_{1,00}/(2Q^\U{tot}_{1}). 
\end{eqnarray}
Here the probabilities are expressed as 
\begin{eqnarray}
p^{(1,1)}_{1,00}&=&
T^2f_2+2T(1-T)f_3+(1-T)^2f_4,\\
p^{(1,1)}_{1,01}&=&
T^2f_1+2T(1-T)f_3+(1-T)^2f_4,\\
p^{(1,1)}_{1,12}&=&
T^2f_5+2T(1-T )f_3+(1-T)^2f_4,\\
p^{(1,2)}_{1,00}&=&(1-T)(T^2f_2+T(1-T)f_3+p^{(1,1)}_{1,00}),\\
p^{(1,2)}_{1,01}&=&(1-T)(T^2f_1+T(1-T)f_3+p^{(1,1)}_{1,01}),\\
p^{(1,2)}_{1,12}&=&
(1-T)(T^2f_5+T(1-T)f_3+p^{(1,1)}_{1,12}),\\
q_{1,00}&=&
p^2_{0}(T\mu)f_4+2p_0(T\mu)p_1(T\mu)f_3+p_1^2(T\mu)f_2,
\\
q_{1,01}&=&p_0^2(T\mu)f_4+2p_{0}(T\mu)p_{1}(T\mu)f_3+p_{1}^2(T\mu)f_1,
\\
q_{1,12}&=&p_0^2(T\mu)f_4+2p_{0}(T\mu)p_{1}(T\mu)f_3+p_{1}^2(T\mu)f_5.
\end{eqnarray}

For Type2, we have 
\begin{eqnarray}
Q^{(1,1)}_{2}&=&p_1^2(\mu)(p^{(1,1)}_{2,00}+p^{(1,1)}_{2,01})/4, \\
e^{(1,1)}\u{2,bit}&=&p_1^2(\mu)p^{(1,1)}_{2,01}/(4Q^{(1,1)}_{2}),\\
Q^{(1,2)}_{2}&=&p_1(\mu)p_{2}(\mu)(p^{(1,2)}_{2,00}+p^{(1,2)}_{2,01})/4,\\
e^{(1,2)}\u{2,bit}&=&p_1(\mu)p_{2}(\mu)p^{(1,2)}_{2,01}/(4Q^{(1,2)}_{2}),\\
Q^\U{tot}_{2}&=&(q_{2,00}+q_{2,01})/4,\\ 
e^\U{tot}_{2}&=&q_{2,01}/(4Q^\U{tot}_{2}), 
\end{eqnarray}
where 
\begin{eqnarray}
p^{(1,1)}_{2,00}&=&p^{(1,1)}_{1,01},\\
p^{(1,1)}_{2,01}&=&p^{(1,1)}_{1,00},\\
p^{(1,2)}_{2,00}&=&p^{(1,2)}_{1,01},\\
p^{(1,2)}_{2,01}&=&p^{(1,2)}_{1,00},\\
q_{2,00}&=&q_{1,01},\\
q_{2,01}&=&q_{1,00}. 
\end{eqnarray}

\subsection{Case (ii) Alice and Bob uses the heralded single photon sources.}

From Ref.~\cite{Qn}, the probability distribution function of the thermal state 
conditioned that the detector $\U{D}_0$ clicked in Fig.~2(b) is 
\begin{eqnarray}
P_n=\frac{1}{P\u{click}}\frac{\mu^n(1-(1-\eta)^n+d)}{(1+\mu)^{n+1}}, 
\end{eqnarray}
where $P\u{click}$ is the probability that the detector $\U{D}_0$ clicks, 
which is described by 
\begin{eqnarray}
P\u{click}=\frac{(1+d)(1+\mu\eta)-1}{1+\mu\eta}. 
\end{eqnarray}
By defining $\eta\u{in}=\eta T$, 
the probability that $n$ photons exist before the BS 
conditioned on the click of $\U{D}_0$ is
\begin{align}
Q_n=\frac{1}{P\u{click}}
\left(
\frac{(1+d)(\mu\eta\u{in})^n}{(1+\mu\eta\u{in})^{n+1}}
-\frac{(\mu\eta\u{in}(1-\eta))^n}
{(1+\mu(\eta\u{in}+\eta-\eta\u{in}\eta))^{n+1}}
\right).
\end{align}

For Type1, 
the relevant equations for $Q_1^{(n,m)}$ and $e^{(n,m)}_{1,\U{bit}}$ 
are expressed by 
\begin{eqnarray}
Q^{(1,1)}_{1}&=&P_1^2(2p^{(1,1)}_{1,00}+p^{(1,1)}_{1,01}+p^{(1,1)}_{1,12})/4,\\
e^{(1,1)}_{1}&=&P_1^2p^{(1,1)}_{1,00}/(2Q^{(1,1)}_{1}),\\
Q^{(1,2)}_{1}&=&P_1P_{2}
(2p^{(1,2)}_{1,00}+p^{(1,2)}_{1,01}+p^{(1,2)}_{1,12})/4,\\
e^{(1,2)}_{1}&=&P_1P_{2}p^{(1,2)}_{1,00}/(2Q^{(1,2)}_{1}),\\
Q^\U{tot}_{1}&=&(2q_{1,00}+q_{1,01}+q_{1,12})/4,\\
e^\U{tot}_{1}&=&q_{1,00}/(2Q^\U{tot}_{1}), 
\end{eqnarray}
where 
\begin{eqnarray}
p^{(1,1)}_{1,00}&=&
\eta\u{in}^2g_4+2\eta\u{in}(1-\eta\u{in})g_2+(1-\eta\u{in})^2g_1,\\
p^{(1,1)}_{1,01}&=&
\eta\u{in}^2g_3+2\eta\u{in}(1-\eta\u{in})g_2+(1-\eta\u{in})^2g_1,\\
p^{(1,1)}_{1,12}&=&
\eta\u{in}^2g_8+2\eta\u{in}(1-\eta\u{in})g_2+(1-\eta\u{in})^2g_1,\\
p^{(1,2)}_{1,00}&=&\eta\u{in}^3g_6
+2\eta\u{in}^2(1-\eta\u{in})g_4+\eta\u{in}^2(1-\eta\u{in})g_7
+3\eta\u{in}(1-\eta\u{in})^2g_2+(1-\eta\u{in})^3g_1,\\
p^{(1,2)}_{1,01}&=&\eta\u{in}^3g_5
+2\eta\u{in}^2(1-\eta\u{in})g_3+\eta\u{in}^2(1-\eta\u{in})g_7
+3\eta\u{in}(1-\eta\u{in})^2g_2+(1-\eta\u{in})^3g_1,\\
p^{(1,2)}_{1,12}&=&\eta\u{in}^3g_9
+2\eta\u{in}^2(1-\eta\u{in})g_8+\eta\u{in}^2(1-\eta\u{in})g_7
+3\eta\u{in}(1-\eta\u{in})^2g_2+(1-\eta\u{in})^3g_1,\\
q_{1,00}&=&
Q_0^2g_1+2Q_0Q_1g_2+Q_1^2g_4+2Q_0Q_2g_7
+2Q_1Q_2g_6+\sum_{n,m=2}^{\infty}Q_nQ_m
\label{eq:pes}\\
q_{1,01}&=&
Q_0^2g_1+2Q_0Q_1g_2+Q_1^2g_3
+2Q_0Q_2g_7+2Q_1Q_2g_5\\
q_{1,12}&=&
Q_0^2g_1+2Q_0Q_1g_2+Q_1^2g_8
+2Q_0Q_2g_7+2Q_1Q_2g_9.
\end{eqnarray}
We note that in equation~(\ref{eq:pes}), we took the pessimistic scenario 
that all of the events for $n\geq 2$ and $m\geq 2$ are detected as the bit error. 
Here $g_1,\ldots ,g_9$ are given by 
\begin{eqnarray}
g_1&=&(1-d)^22d^2,\\
g_2&=&(1-d)^2d,\\
g_3&=&(1-d)^2(2a^2b^2+(a^4+b^4)d),\\
g_4&=&g_3-(1-d)^22a^2b^2,\\
g_5&=&(1-d)^2(9(a^4b^2+a^2b^4)+3(a^6+b^6)d)/4,
\\
g_6&=&(1-d)^2((a^4b^2+a^2b^4)+3(a^6+b^6)d)/4,
\\
g_7&=&g_3/2,\\
g_8&=&(1-d)^2 2a^2b^2 (1+d),\\
g_9&=&(1-d)^2(a^6+b^6+3(a^4b^2+a^2b^4)d)/4.
\end{eqnarray}

For Type2, we have 
\begin{eqnarray}
Q^{(1,1)}_{2}&=&P_1^2(p^{(1,1)}_{2,00}+p^{(1,1)}_{2,01})/4, \\
e^{(1,1)}_{2}&=&P_1^2p^{(1,1)}_{2,01}/(4Q^{(1,1)}_{2}),\\
Q^{(1,2)}_{2}&=&P_1P_{2}(p^{(1,2)}_{2,00}+p^{(1,2)}_{2,01})/4,\\
e^{(1,2)}_{2}&=&P_1P_{2}p^{(1,2)}_{2,01}/(4Q^{(1,2)}_{2}),\\
Q^\U{tot}_{2}&=&(q_{2,00}+q_{2,01})/4,\\
e^\U{tot}_{2}&=&q_{2,01}/(4Q^\U{tot}_{2}), 
\end{eqnarray}
where 
\begin{eqnarray}
p^{(1,1)}_{2,00}&=&p^{(1,1)}_{1,01},\\
p^{(1,1)}_{2,01}&=&p^{(1,1)}_{1,00},\\
p^{(1,2)}_{2,00}&=&p^{(1,2)}_{1,01},\\
p^{(1,2)}_{2,01}&=&p^{(1,2)}_{1,00},\\
q_{2,00}&=&q_{1,01},\\
q_{2,01}&=&q_{1,00}. 
\end{eqnarray}

\end{widetext}

\end{document}